\documentstyle[preprint,eqsecnum,aps,epsf]{revtex}
\tightenlines
\begin{document}

\draft
\preprint{IFUM 572/FT, May 1997}

\title{Schwinger-Dyson Equation, Area Law and Chiral Symmetry in QCD}

\author{G. M. Prosperi}

\address{
Dipartimento di Fisica dell'Universit\`{a} -- Milano\\
INFN, Sezione di Milano -- Via Celoria 16, 20133 Milano
}

\maketitle
\begin{abstract}
\indent 
A Schwinger-Dyson equation for the quark propagator is derived in the context 
of a Bethe-Salpeter second order formalism developped in preceding papers and
of the Minimal Area Law model for the evaluation of the Wilson loop. We 
discuss how the equal time straight line 
approximation has to be modified to include correctly trajectories going 
backwards in time. We also show, by an appropriate selection of the 
solution of the SD equation, that in the limit of zero quark mass chiral 
symmetry breaking and a zero mass pseudoscalar meson actually occur. The 
inclusion of backward quark trajectories proves to be essential to make the 
model consistent with Goldstone's theorem.
\end{abstract}



\section{Introduction}

     The problem of the chiral symmetry breaking in QCD \cite{weinberg} and 
that of the consistency with Goldstone's theorem of approximations in the 
kernels of the Schwinger-Dyson (SD) and Bethe-Salpeter (BS) equations 
has been discussed by various authors \cite{delbourgo,nambu}. To my knoledge, 
however, in all such papers only the perturbative part of the kernel was 
actually derived from QCD. To this an {\it ad hoc} infrared singular term 
was added only for the sake of convenience and to make the treatment of the 
resulting equations as easy as possible.

     On the other hand, taking advantage of a four-dimensional path integral
representation for a {\it second order} quark-antiquark Green function 
$ H(x_1, x_2; y_1, y_2) $, in a preceding paper we succeeded in 
obtaining
a BS like equation for such quantity entirely from first principles, a part 
the use of the so called Minimal Area Law (MAL) model for the evaluation of 
the Wilson loop \cite{bmp,BP95}.

     In this paper I first show that along similar lines a Schwinger-Dyson 
equation can be derived for the {\it colorless} second order quark 
propagator $H(x-y)$ occurring in the BS equation. Then I discuss how 
the equal time straight line approximation to the minimal area has to 
be modified in order to make the formalism consistent with Goldstone's 
theorem in the limit of zero quark masses.

     The mentioned path integral representation for $H(x_1,x_2;y_1,y_2)$ 
is in turn a consequence of the similar Feynmann-Schwinger (FS)
representation for a one particle propagator in an external field. 
Its interest lies in the fact that the gauge field appears in it only 
in terms of the Wilson loop made by the quark and the antiquark 
trajectories (connecting $y_1$ to $x_1$ and $y_2$ to $x_2$) closed by 
two Schwinger strings \cite{peskin,BP95,bmp},

 \begin{equation}
W = {1\over 3}
   \langle {\rm Tr} {\rm P} \exp \{ i g  \oint_{\Gamma_{\bar q q}} dx^{\mu} 
   A_{\mu} \}  \rangle  \ .
\label{eq:loop}
  \end{equation}

\noindent 
As usual in (\ref{eq:loop})  $A_\mu = {1\over 2} A_\mu^a 
\lambda^a$ is a colour matrix, ${\rm P}$  the ordering prescription along 
the loop and the expectation value stands for a  functional 
integration  on the gauge field alone.  The MAL model consists
in assuming that the logaritm of the Wilson loop can be written as the 
sum of a perturbative and an area term

  \begin{equation}
i \ln W = i (\ln W)_{\rm pert} + \sigma S_{\rm min}   \,  ,
\label{eq:wl}
   \end{equation}

\noindent
$\sigma$ being the so called string tension and $S_{\rm min}$ 
the minimal area enclosed by $\Gamma_{\bar q q}$. In practice 
$S_{\rm min}$ is replaced by its equal time straight line approximation,
consisting in setting

\begin{equation}
S_{\rm min}= \int_0^{s_1}
 d\tau_1 \int_0^{s_2} d\tau_2 \delta(z_{10}-z_{20})
  \vert {\bf z}_1 -{\bf z}_2\vert
\int_0^1 d\lambda
 \Big \{ {\dot{z}_{10}^{ 2}} {\dot{z}_{20}^{ 2}} -
 (\lambda \dot{\vec z}_{1{\rm T}} \dot{z}_{20}
 + (1-\lambda) \dot{\vec z}_{2 {\rm T}}
 \dot{z}_{10} )^2 \Big \}^{1\over 2} \ ,
\label{eq:lstrai}
\end{equation}

\noindent
where $ z_1=z_1(\tau_1) $ and  $ z_2=z_2(\tau_2) $ (with 
$ 0\le \tau_1 \le s_1 $
and  $0\le \tau_2 \le s_2$) are the parametric equations for the quark and 
the antiquark trajectories respectively and $\dot{\vec z}_{1 {\rm T}} $, 
$\dot{\vec z}_{2 {\rm T}}$ stand for the transverse  components of 
$ \dot {\vec z}_1$ and $ \dot {\vec z}_2 $ [$\dot{z}_{j {\rm T}}^h 
= (\delta^{hk} -\hat{r}^h \hat{r}^k ) \dot{z}_j^h $ with 
 ${\vec r} = {\bf z}_1 -{\vec z}_2 $ and $\hat{r}^h = {r^h \over r}$].

    In the limits $ x_2 \rightarrow x_1 $ and  $ y_2 \rightarrow y_1 $
(i.e. when the initial and respectively the final points of the quark 
and the antiquark trajectories coincide), Eq. (\ref{eq:lstrai}) is
correct only up to the second order terms in $\dot z_1$ and $\dot z_2$ 
in the general case, but it
becomes exact in two different significant situations: when the two 
trajectories lie on a plane (with one time and one space dimension) and
when they form a double elicoid. This fact may justify its use even in 
a fully relativistic case.

    Notice however that the right hand side of (\ref{eq:lstrai}) is of a 
purely $q \bar q$ interaction type, while 
the perturbative contribution contains both an $q \bar q$ interaction 
term and two corresponding $q$ and $\bar q$ selfinteraction terms. As a 
consequence of this lack of symmetry, in the limit of zero quark mass
(\ref{eq:lstrai}) proves to be inconsistent with the Goldstone theorem,
if dynamical symmetry breaking occurs. We shall see that we obtain 
consistency if we replace (\ref{eq:lstrai}) with

\begin{eqnarray}
S _{\rm min}&=& \int_0^{s_1}
   d\tau_1  \int_0^{s_2} d\tau_2 \delta(z_{10}-z_{20})
   \vert {\vec z}_1 -{\vec z}_2\vert
   \epsilon (\dot{z}_{10})
   \epsilon (\dot{z}_{20}) \nonumber \\
  & & \qquad \qquad \qquad \times \int_0^1 d\lambda
   \Big \{ {\dot{z}_{10}^{ 2}} {\dot{z}_{20}^{ 2}} -
   (\lambda \dot{\vec z}_{1{\rm T}} \dot{z}_{20}
   + (1-\lambda) \dot{\vec z}_{2 {\rm T}}
   \dot{z}_{10} )^2 \Big \}^{1\over 2}  - \nonumber \\
 & & - \int_0^{s_1}
   d\tau_1 \int_0^{\tau_1} d\tau_1^{\prime} \delta(z_{10}-z_{10} ^{\prime})
  \vert {\vec z}_1 -{\vec z}_1 ^{\prime}\vert
   \epsilon (\dot{z}_{10})
   \epsilon (\dot{z}_{10}^{\prime} ) \nonumber \\
 & & \qquad \qquad \qquad \times \int_0^1 d\lambda
   \Big \{ {\dot{z}_{10}^{ 2}} {\dot{z}_{10}^{\prime 2}} -
   (\lambda \dot{\vec z}_{1{\rm T}} \dot{z}_{10} ^{\prime}
   + (1-\lambda) \dot{\vec z}_{1 {\rm T}} ^{\prime}
   \dot{z}_{10} )^2 \Big \}^{1\over 2} - \nonumber\\
& &  -\int_0^{s_2}
    d\tau_2 \int_0^{\tau _2} d\tau_2 ^{\prime} \delta(z_{20}-z_{20} ^{\prime})
    \vert {\vec z}_2 -{\vec z}_2 ^{\prime}\vert
   \epsilon (\dot{z}_{20})
   \epsilon (\dot{z}_{20}^{\prime}) \nonumber \\
& &  \qquad \qquad \qquad \times \int_0^1 d\lambda
   \Big \{ {\dot{z}_{20}^{ 2}} {\dot{z}_{20}^{\prime 2}} -
   (\lambda \dot{\vec z}_{2{\rm T}} \dot{z}_{20} ^{\prime}
    + (1-\lambda) \dot{\vec z}_{2 {\rm T}} ^{\prime}
   \dot{z}_{20} )^2 \Big \}^{1\over 2} \, , \nonumber\\
\label{eq:nstrai}
\end{eqnarray}

\noindent
where we have used $z_1 ^{\prime}$ and  $z_2 ^{\prime}$ for  $z_1 (\tau_1
^{\prime})$ and  $z_2 (\tau_2 ^{\prime})$ respectively and $\epsilon (t)$ 
for the sign function.

    In fact (\ref{eq:lstrai}) breaks down necessarily if the trajectories 
can go backwards in time. On the contrary, if they never do that, 
(\ref{eq:nstrai}) becomes identical to (\ref{eq:lstrai})  (since in this case 
the first term is positive and the other two vanish), but for a plane or an 
elicoid loop it remains exact, as we shall see, even if they do. The correct 
inclusion of such more general trajectories turns out to be essential and 
affects substantially the SD equation.

    The resulting SD equation turns out to be equivalent to a system of four 
non linear integral equations envolving four scalar quantities, $h_0(k),\dots
h_3(k)$, which appear as coefficients of product of Dirac matrices with 
different tensorial character in the expression of the quark propagator. 
We are not able to handle rigorously such a system. However, its struture
suggests the possibility of four different classes of solutions corresponding
to different sets of $h_s(k)$ being indentically zero. Actually, if we 
assume a smooth behaviour as the quark masses go to zero,  we can argue
that the first class, corresponding to a purely scalar propagator ($h_1=h_2=h_3
\equiv 0$), is empty. In fact, if one tries to construct such a solution by 
iteration, starting from the free propagator, the occurrence in the 
above limit of the strong infrared singularity due to the area law term 
(\ref{eq:nstrai}) is found to prevent the procedure to converge.

    As an elimination criterion of other classes one can use the unexpected 
occurrence of zero mass bound states in the corresponding BS equations 
and the consistency with Goldstone's theorem. Indeed in field theory the 
occurence of zero mass bound states appears as an exceptional circustance,
expected only in connection with a spontaneously broken symmetry. In this
way also the second ($h_3 \equiv 0$) and the third classes ($h_1=h_2 
\equiv 0$) can be rejected as spurious and we are left with the fourth 
class alone, for which $ h_0, \dots h_3 $ are all different from zero. 
Inside this fourth class various possible solutions can still be
considered, which reduce to the above classes when the quark masses
$m_1$ and $m_2$ vanish. By similar arguments also such solutions can 
be eliminated but one, that for which $h_1$ and $h_2$ keep 
different from zero, while $h_3$ vanishes for $m_1,m_2 \to 0$. 

    The last solution corresponds to a breaking of the chiral symmetry in 
the vanishing quark mass limit and to the occurrence of a zero mass 
pseudoscalar solution of the BS equation, as one expects. Notice that what 
rules out the possibility of a solution of the first class, or reducing to
the first class for $m_1,m_2 \to 0$, is the already mentioned infrared 
singularity. Therefore this is ultimately responsible for the chiral 
symmetry breaking.

   The plan of the remaining part of the paper is the following: In 
Sec. II, besides the first and the second order $q \bar q$ Green functions 
$ G^{\rm gi}(x_1, x_2; y_1, y_2)$ and $ H^{\rm gi}(x_1, x_2; y_1, y_2)$, 
already considered in \cite{bmp}, we introduce the corresponding single 
quark propagators $ G^{\rm gi}(x-y) $ and $H^{\rm gi}(x-y) $. In Sec. III
we discuss Eq. (\ref{eq:nstrai}); we also introduce the functions 
$ H(x_1, x_2; y_1, y_2)$ and $ H(x-y)$, which are obtained by neglecting
certain contributions related to the Schwinger strings, but become 
identical to $ H^{\rm gi}(x_1, x_2; y_1, y_2)$ and $ H^{\rm gi}(x-y) $ 
in the limits $ x_2 \rightarrow x_1 $ and  $ y_2 \rightarrow y_1 $ 
(notice that only such limits are relevants for bound states and chiral
symmetry breaking). In Sec.IV we derive a SD equation for $ H(x-y) $ 
along lines similar to those followed to obtain the BS equation for 
$ H(x_1, x_2; y_1, y_2)$ in \cite{bmp}. In Sec. V we rewrite the BS 
equation in a form which makes it easier to compare to the SD equation.
In Sec. VI we discuss the possible solutions of the SD equation and, in the 
limit of vanishing quark masses, the chiral symmetry breaking and the BS 
equation for a zero mass bound state. Finally in 
Sec, VII we summarize the results and make some additional remarks.

   Notice that no attempt of a numerical explicit resolution of the
resulting SD equation is made in this paper.


\section{Green functions and Feynmann-Schwinger representations}

     The quark-antiquark and the single quark gauge invariant Green 
functions are defined as 

\begin{eqnarray}
G^{\rm gi}(x_1,x_2,y_1,y_2) &=&
\frac{1}{3}\langle0|{\rm T}\psi_2^c(x_2)U(x_2,x_1)\psi_1(x_1)
\overline{\psi}_1(y_1)U(y_1,y_2)  \overline{\psi}_2^c(y_2)
|0\rangle =
\nonumber\\
&=& \frac{1}{3} {\rm Tr_C} \langle U(x_2,x_1)
 S_1(x_1,y_1;A) U(y_1,y_2)
\tilde S_2(y_2,x_2;-\tilde A) \rangle
\label{eq:qqgauginv}
\end{eqnarray}

\noindent
and

\begin{eqnarray}
G^{\rm gi}(x-y) &=&
\langle0|{\rm T}U(y,x)\psi(x)
\overline{\psi}(y)
|0\rangle =
\nonumber\\
&=& i {\rm Tr_C} \langle U(y,x)
 S(x,y;A) \rangle \, ,
\label{eq:qgauginv}
\end{eqnarray}

\noindent
where $\psi^c$ denotes the charge-conjugate fields, the tilde and 
${\rm Tr_C}$ the transposition and the trace respectively over the 
color indices alone, $U$ the path-ordered gauge string (Schwinger 
string)

\begin{equation}
U(b,a)= {\rm P}  \exp \left \{ig\int_a^b dx^{\mu} \, A_{\mu}(x) 
\right \}
\label{eq:col}
\end{equation}

\noindent
(the integration being along the straight line joining $a$ to $b$, 
although otherwise specified, and P being the ordering operator on 
the color matrices), $S$, $S_1$ and $S_2$ the quark propagators in the 
external gauge field $A^{\mu}$ and obviously
 
\begin{equation}
\langle f[A] \rangle = \frac{\int {\cal D}[A] M_f(A)
f[A] e^{iS[A]}}
{\int {\cal D}[A] M_f(A) e^{iS[A]}} \> ,
\label{eq:med}
\end{equation}

\noindent
$S[A]$ being the pure gauge field action and $M_f(A)$ the determinant
resulting from the explicit integration on the fermionic fields (in 
practice however $M_f(A)=1$ in the adopted approximation). Notice that 
we shall suppress indices specifying the quarks as a rule when dealing 
with single quark quantities.

    The  propagator  $S$ is supposed to be defined 
by the equation 

\begin{equation}
( i\gamma^\mu D_\mu -m) S(x,y;A) =\delta^4(x-y)
\label{eq:propdir}
\end{equation}
 
\noindent
and the appropriate  boundary conditions.

  As in \cite{bmp} we introduce the second order propagator 

\begin{equation}
(D_\mu D^\mu +m^2 -{1\over 2} g \, \sigma^{\mu \nu} F_{\mu \nu})
\Delta^\sigma (x,y;A) = -\delta^4(x-y)
\label{eq:propk}
\end{equation}

\noindent
and write

\begin{equation} 
S(x,y;A) = (i \gamma^\nu D_\nu + m) \Delta^\sigma(x,y;A) ,
\label{eq:fsord}
\end{equation}

\noindent
with ($\sigma^{\mu \nu} = {i\over 2} [\gamma^\mu, \gamma^\nu]$). Then,
after replacing (\ref{eq:fsord}) in (\ref{eq:qgauginv}) and 
(\ref{eq:qqgauginv}), using an appropriate derivative it is possible
to take the differential operator out of the angle brackets and write

\begin{equation}
G^{\rm gi}(x_1,x_2; y_1,y_2) =-(i \gamma_1^\mu \bar{\partial}_{1 \mu} 
 + m_1) ( i \gamma_2^\nu \bar{\partial}_{2 \nu} +m_2) 
H^{\rm gi}(x_1,x_2;y_1,y_2)
\label{eq:qqeqg}
\end{equation}

\noindent
and

\begin{equation}
G^{\rm gi}(x - y) = (i \gamma_1^\mu \bar{\partial}_{\mu} 
 + m) H^{\rm gi}(x-y)
\label{eq:qeqg}
\end{equation}

\noindent
with

\begin{equation}
H^{\rm gi}(x_1,x_2;y_1,y_2) = -{1\over 3} {\rm Tr _C}
\langle U(x_2,x_1) \Delta_1^\sigma (x_1,y_1;A)
 U(y_1,y_2) \tilde{\Delta}_2^\sigma (x_2,y_2;-\tilde{A})\rangle .
\label{eq:qqeqh}
\end{equation}

\noindent
and

\begin{equation}
H^{\rm gi}(x-y) = i {\rm Tr _C}
\langle U(y,x) \Delta ^\sigma (x,y;A) \rangle .
\label{eq:qeqh}
\end{equation}

In the last equations $\bar{\partial}_{\mu}$ stands for the ordinary 
derivative when acting on the propagators, while it refers to an 
appropriate distortion of the path when acting on a Schwinger string.
In more explicit terms one can also write

\begin{equation}
\bar{\partial}_{b^\mu} U(b,a)={\partial}_{b^\mu} U(b,a)- \int_0^1 d \lambda
     \, \lambda (b^\rho - a^\rho) 
     {\delta \over \delta S^{\mu \rho}(a+ \lambda (b-a))} U(b,a)
\label{eq:modfder}
\end{equation}

\noindent
(with $ \delta S^{\mu \nu} = dz^\mu \delta z^\nu - dz^\nu \delta z^\mu 
$ and the functional derivative being defined through an arbitrary 
deformation, $ z \rightarrow z + \delta z $, of the line connecting 
$a$ to $b$ starting from the straight line) and a similar expression 
for $\bar \partial_{a^ \mu}$.
\footnote{Given a functional $\Phi[\gamma_{ab}]$
 of the curve $\gamma_{ab}$ with ends $a$ and $b$, let us assume that the 
 variation of $\Phi$ consequent 
 to an infinitesimal  modification  of the curve $\gamma \rightarrow \gamma
 +\delta \gamma $ can be expressed as the sum of various terms
 proportional respectively to $\delta a$, to $\delta b$ and to the single
 elements $\delta S^{\rho \sigma}(x)$ of the surface swept by the curve.
 Then, the derivatives $ \bar{\partial}_{a^{\rho}}$,
 $ \bar{\partial}_{b^{\rho}}$ and $\delta / \delta 
 S^{\rho \sigma}(x)$ are defined by the 
 equation $ \delta \Phi = \delta a^{\rho}
 \bar{\partial}_{a^\rho} \Phi 
 + \delta b^{\rho} \bar{\partial}_{b^\rho} \Phi   + {1 \over 2}\int_{\gamma} 
 \delta S^{\rho \sigma}(x) \delta  \Phi / \delta S^{\rho \sigma}(x)$. 
 For a Schwinger string we have $ \delta U(b,a) = \delta b^\rho ig A_\rho (b)
 U(b,a) - \delta a^\rho U(b,a) ig A_\rho (a) + {ig \over 2} \int_a^b 
 \delta S^{\rho \sigma}
 (z) {\rm P}(-F_{\rho \sigma} (z) U(b,a) $ and so
 $\bar{\partial}_{a^{\rho}} U = -ig U A_{\rho}(a) $,
 $\bar{\partial}_{b^{\rho}} U = ig A_{\rho}(b) U $ and $ {\delta \over 
 \delta S^{\rho \sigma}(z)} U = {\rm P}[-ig F_{\rho \sigma} (z) U] $.}

  For the second order propagator we have the Feynman-Schwinger 
representation \cite{bmp}

\begin{equation}
\Delta^\sigma (x,y; A)= -{i \over 2} \int_0^\infty ds \int_y^x
   {\cal D} z \exp [-i \int_0^s d\tau {1\over 2} (m^2 
   +\dot z ^2)] {\cal S}_0^s {\rm P} \exp[ ig \int_0^s d \tau
   \dot z ^\mu A_{\mu}(z)] \, , 
\label{eq:partbis}
\end{equation}

\noindent
with 

\begin{equation}
{\cal S}_0^s = {\rm T} \exp \Big [ -{1\over 4} \int_0^s d \tau 
 \sigma^{\mu \nu} {\delta \over \delta S^{\mu \nu}(z)}
\Big ]\, .
\label{eq:defop}
\end{equation}

$T$ and $P$ being the ordering prescriptions along the path acting on 
the spin and on the color matrices respectively. Replacing 
(\ref{eq:defop}) in (\ref{eq:qqeqh}) and (\ref{eq:qeqh}) we obtain

\begin{eqnarray}
H^{\bf gi}(x_1,x_2;y_1,y_2) & &= ({1 \over 2})^2 \int_0^{\infty} d s_1
\int_0^{\infty} d s_2
 \int_{y_1}^{x_1} {\cal D} z_1\int_{y_2}^{x_2} {\cal D} z_2
 \nonumber \\
& & \exp \Big \{  -{i \over 2} 
 \int_0^{s_1} d\tau_1 (m_1^2 +\dot{z}_1^2) - {i\over 2} \int_0^{s_2}
d\tau_2 (m_2^2 +\dot{z}_2^2)\big \} \nonumber \\
& &  \times {\cal S}_0^{s_1} {\cal S}_0^{s_2}
{1\over 3} \langle {\rm Tr} \, {\rm P}
 \exp  \big \{ ig \oint_{\Gamma_{\bar q q}} dz^\mu A_{\mu} 
(z)   \} 
\rangle    ,
\label{eq:hqq}
\end{eqnarray}

\noindent
and

\begin{eqnarray}
H^{\bf gi}(x-y) & &= {1 \over 2} \int_0^{\infty} d s
 \int_y^x {\cal D} z
 \exp \big \{  -{i \over 2} 
 \int_0^{s} d\tau (m^2 +\dot{z}^2) \nonumber \\
& &  \times {\cal S}_0^s
 \langle {\rm Tr}\, {\rm P}
 \exp  \big \{ ig \oint_{\Gamma _q} dz^\mu A_{\mu} 
(z)   \} 
\rangle    .
\label{eq:hq}
\end{eqnarray}

\noindent
Here, the loop $\Gamma_{\bar q q}$ occuring in the 4-points function is 
made by the quark world line $\gamma_1$, the antiquark world line 
$\gamma_2$ followed in the reverse direction, and the two straight 
lines $x_1 x_2$ and $y_2 y_1$ (Fig. \ \ref{fig1}), as we already mentioned,
on the contrary the loop $\Gamma _q$ occurring in the 2-points function
is made simply by the quark trajectory $\gamma$ connecting $y$ to $x$ 
and by the straight line $yx$ (Fig. \ \ref{fig2}).


\section{Wilson loop with backward trajectories}

\indent 
     Now we want to apply Eq. (\ref{eq:wl}) to the evaluation of the Wilson 
loop integral occurring both in (\ref{eq:hqq}) and in (\ref{eq:hq}). 

     At the lowest order the perturbative term can be written

\begin{equation}
i (\ln W)_{\rm pert} = -{2 \over 3} g^2 \oint dz^\mu \oint dz^{\nu \prime}
D_{\mu \nu}(z-z^\prime)
\label {eq:perta}
\end{equation}

\noindent
$ D_{\mu \nu}(z-z^\prime) $ being the free gauge propagator, which can be 
grafically represented by a waving line connecting the point $z$ and 
$ z^{\prime} $ on the loop. If we neglect the contribution coming from
lines connecting a point on a trajectory to a point on a string or two
point on the strings, we can write for $\Gamma_{\bar q q}$

\begin{eqnarray}
& &i (\ln W)_{\rm pert} = {4\over 3} g^2 \int_0^{s_1} d \tau_1
\int_0^{s_2} d\tau_2 D_{\mu \nu}(z_1-z_2)
 \dot{z}_1^{\mu} 
\dot{z}_2^{\nu}- \nonumber \\
& & -{4\over 3} g^2 \int_0^{s_1} d \tau_1
\int_0^{\tau _1} d\tau_1^{\prime}
 D_{\mu \nu}(z_1-z_1^{\prime}) \dot{z}_1^{\mu} 
\dot{z}_1^{\prime\nu}-
{4\over 3} g^2 \int_0^{s_2} d \tau_2
\int_0^{\tau _2} d\tau_2^{\prime} D_{\mu 
\nu}(z_2-z_2^{\prime}) \dot{z}_2^{\mu} 
\dot{z}_2^{\prime \nu}+ \dots
\label{eq:pertb}
\end{eqnarray}

\noindent
and for $\Gamma_q$

\begin{equation}
i (\ln W)_{\rm pert} =  -{4\over 3} g^2 \int_0^{s} d \tau
   \int_0^{\tau } d\tau^{\prime}
   D_{\mu \nu}(z-z^{\prime}) \dot{z}^{\mu} 
   \dot{z}^{\prime \nu}+ \dots
\label{eq:pertc}
\end{equation}

    Obviously the above approximation would not make sense in general,
but we are eventually interested in the two limits $ y \to x $
and $ y_2 \to x_2 $, $ y_1 \to x_1 $ and in which Eq.'s 
(\ref{eq:pertc}) and (\ref{eq:pertb}) become exact.

    Let us pass to consider the area term in (\ref{eq:wl}) and refer to 
the loop $\Gamma_{\bar q q}$ first.

    In general one can writes

\begin{equation}
S_{\rm min} = \int_{t_{i}}^{t_{f}} dt
 \int_0^1 d\lambda
 \Big [ -({\partial u^\mu \over \partial t} {\partial u_\mu 
\over \partial t })({\partial u^\nu \over \partial \lambda} 
{\partial u_\nu \over \partial \lambda })+
 ({\partial u^\mu \over \partial t} {\partial u_\mu 
\over \partial \lambda})^2 \Big ]^{1\over 2} \, ,
\label{eq:smin}
\end{equation}

\noindent
$x^\mu = u^\mu (\lambda,t) $ being  the equation of the minimal surface
with  contour $\Gamma_{\bar q q}$. Since (\ref{eq:smin}) is invariant under 
reparametrization, a priori the parameter  $t$  can be everything.
So we can assume  that for fixed $t$ one has
 
\begin{equation}
u^\mu(1,t)= z_1^\mu(\tau_1(t)),\quad \quad \quad
u^\mu(0,t)= z_2^\mu(\tau_2(t)).
\label{eq:udef}
\end{equation}

\noindent
However, if $\gamma_1$ and $\gamma_2$ never go backwards in time, 
we can also assume $t$ to be the ordinary time,  $u^0(s,t)\equiv t$.
Then $\tau_1 (t)$ and $\tau_2(t)$ are specified by the equation

\begin{equation}
z_1^0(\tau_1) =z_2^0 (\tau_2)= t \, .
\label{eq:zzero}
\end{equation}

\noindent
and the equal time straight line approximation consits in setting

\begin{equation}
u^0(\lambda, t) = t \, , \quad \quad \quad u^k(\lambda, t) 
= \lambda z_1^k (\tau_1(t)) + (1-\lambda)z_2^k(\tau_2(t))
\label{eq:stl}\, .
\end{equation}

\noindent
Replacing this in (\ref{eq:smin}) and performing a change of 
integration variables, one obtains (\ref{eq:lstrai}). 

  As regards to Eq. (\ref{eq:nstrai}), notice first that, as we already 
mentioned, this equation is identical to (\ref{eq:lstrai}), if the 
trajectories never go backwards in time. In order to discuss the 
situation we have when this is not the case, we assume that the extreme
points coincide ($x_1=x_2$ and $y_1=y_2$) and the trajectories lie on
a plane. Then let us refer to Fig. \ \ref{fig3} and consider first the
interaction term. For $z_2$ between the points $R$ and $S$ we have 
three different values of $\tau_1$ which solve (\ref{eq:zzero}) and 
therefore three different intervals of $\tau_1$ which give
contribution; these correspond to $z_1$ between $A$ and $B$ or between 
$B$ and $C$ or between $C$ and $D$. In absolute value such contributions
equal respectively the areas RABSR, RCBSR and 
RCDSR. However, when attention is paid to the sign factors appearing in
front of the integral in $\lambda$ in (\ref{eq:nstrai}), one realizes
that the second contribution is negative, while the other two are 
positive. In this case the algebraic sum of the three areas equals
simply the sum of CABC and RCDSR, and this amounts to the total area
enclosed in the loop in the strip between the lines RA and SB. 

    In a similar way, we can see that the contribution to the above strip 
coming from the self-energy term of particle 1 is also made by the sum of
three terms (corresponding to $z_1^\prime$ between $A$ an $B$  and $z_1$ 
betwen $B$ and $C$ or $z_1^\prime$ between $A$ and $B$ and $z_1$ between 
$C$ and $D$ or $z_1^\prime$ between $B$ and $C$ and $z_1$ between $C$ and
$D$), but in this case such sum is zero. 

    An identical situation occurs for the strip between LE and $x{\rm 
F}$. On the contrary, it is clear that no contribution comes from the 
interaction term to the area HFGH corresponding to a time larger than 
$x_1^0=x_2^0$, while this area
is fully taken into account by the self-energy term. 

   In conclusion, for $\Gamma_{\bar q q}$ on a plane and with the 
extremes of  $\gamma_1$ and  $\gamma_2$ coinciding, the Eq. 
(\ref{eq:nstrai}) is exact. For reason of symmetry the same must be true
for a $\Gamma_{\bar q q}$ obtained from the preceding one
by screwing the trajectories to elics along a line parallel to the time 
axis. In a general case (\ref{eq:nstrai}) can be only an approximation,
but we expect to be a good one, since the contribution coming from
trajectories very far from the above classes should be strongly suppressed
in the path integral (being far also from the minimun of the exponent).

   In a similar way in the case of $\Gamma _q$ we shall set 

\begin{eqnarray}
S_{\rm min}&=&  -\int_0^{s}
 d\tau \int_0^{\tau} d\tau ^{\prime} \delta(z_0-z _0 ^ \prime)
  \vert {\vec z} -{\vec z} ^{\prime}\vert
\epsilon (\dot{z}_0)
\epsilon (\dot{z}_0  ^\prime)
\int_0^1 d\lambda
 \Big \{ {\dot{z}_0^{ 2}} {\dot{z}_{0}^{\prime 2}}-  \nonumber\\
 &&- (\lambda \dot{\vec z}_{\rm T} \dot{z}_{0} ^{\prime}
 + (1-\lambda) \dot{\vec z}_{\rm T} ^{\prime}
 \dot{z}_{0} )^2 \Big \}^{1\over 2} 
\label{eq:rstrai}
\end{eqnarray}

\noindent
and notice that even this equation becomes exact for $y$ = $x$ if  
$\Gamma_q$ is of one of the types discussed above.

     As we shall see, for zero quark masses (\ref{eq:nstrai}) and 
(\ref{eq:rstrai}), taken together, are approximations consistent with 
chiral symmetry breaking and Goldstone theorem.

    Replacing (\ref{eq:pertb}) and (\ref{eq:nstrai}) in (\ref{eq:hqq}) and 
(\ref{eq:pertc}) and (\ref{eq:rstrai}) in (\ref{eq:hq}) we obtain the following 
equations

\begin{eqnarray}
H(x_1,x_2;y_1,y_2) &=& ({1 \over 2})^2 \int_0^{\infty} d s_1
  \int_0^{\infty} d s_2
  \int_{y_1}^{x_1} {\cal D} z_1\int_{y_2}^{x_2} {\cal D} z_2
   \exp \Big \{  -{i \over 2} 
  \sum_{j=1}^2  \int_0^{s_j} d\tau_j (m_j^2 +\dot{z}_j^2) \Big \} \nonumber \\
 & & \times {\cal S}_0^{s_1} {\cal S}_0^{s_2}
 \exp \Big \{ i \sum_{j=1}^2 \int_0^{s_j} d \tau_j  \int_0^{\tau_j} 
  d \tau _j ^\prime 
    E( z_j -z_j^\prime ; \dot z _j , \dot z _j^\prime) - \nonumber\\
& & \qquad \qquad \qquad \qquad -i \int_0^{s_1} d \tau_1  \int_0^{s_2} 
    d \tau _2 
   E( z_1 -z_2 ; \dot z _1 , \dot z _2) \Big \}
\label{eq:fsqq}
\end{eqnarray}

\noindent
and

\begin{eqnarray}
H(x-y) & &= {1 \over 2} \int_0^{\infty} d s
 \int_y^x {\cal D} z
 \exp \big \{  -{i \over 2} 
 \int_0^{s} d\tau (m^2 +\dot{z}^2) \big \} \nonumber \\
& &  \qquad \qquad \qquad \qquad \times {\cal S}_0^s
\exp \big \{i \int_0^s \int_0^\tau 
E(z - z^\prime; \dot z , \dot z ^\prime) \big \} \ ,
\label{eq:fsq}
\end{eqnarray}

\noindent
where we have set

\begin{equation}
E(\zeta;p,p^\prime) = E_{\rm pert} (\zeta;p,p^\prime) + 
      E_{\rm conf} (\zeta;p,p^\prime) 
\label{eq:eea}
\end{equation}

\noindent
with 

\begin{equation}
\left \{
\begin{array}{ll}
E_{\rm pert} & = 4 \pi {4 \over 3} \alpha_{\rm s} D_{\mu \nu}(\zeta)
   p^\mu p^{\prime \nu}  \nonumber \\
E_{\rm conf} & = \delta (\zeta _0) \vert {\vec \zeta} \vert \epsilon (p_0)
   \epsilon (p_0^{\prime})
   \int_0^1 d \lambda \{ p_0^2 p_0^{\prime 2} - [\lambda p_0^{\prime}
   {\vec p}_{\rm T}  
   + (1- \lambda) p_0 {\vec p}_{\rm T} ^{\prime} ]^2 \}^{1 \over 2} 
\label{eq:eeb}
\end{array} \right.
\end{equation}

\indent
    Notice that in principle the quantities $H(x_1, x_2;y_1, y_2)$ and 
$H(x-y)$ as defined by (\ref{eq:fsqq})-(\ref{eq:eeb}) may differ 
significantly from the original $H^{\rm gi}(x_1, x_2;y_1, y_2)$ and
$H^{\rm gi}(x-y)$ for arbitrary arguments, since, in writing such 
equations, we have neglected the contributions to the 
Wison loops coming from the strips betwen $x_1^0$ and $x_2^0$,
$y_1^0$ and $y_2^0$, $y^0$ and $x^0$, for what concerns the area parts, 
and from propagators
string-string or string-trajectory for the perturbative parts. However, 
as we mentioned, the two couples of quantities must concide
in the limits $x_2 \to x_1$, $y_2 \to y_1$, and $y \to x$.

    We can also define the quantities 
   
\begin{equation}
G(x_1,x_2; y_1,y_2) =-(i \gamma_1^\mu {\partial}_{1 \mu} 
 + m_1) ( i \gamma_2^\nu {\partial}_{2 \nu} +m_2) 
H(x_1,x_2;y_1,y_2) \, ,
\label{eq:qqmdfeqg}
\end{equation}

\noindent
and

\begin{equation}
G(x - y) = i(i \gamma^\mu {\partial}_{\mu} 
 + m) H(x-y) \, ,
\label{eq:qmdfeqg}
\end{equation}

\noindent
where we have neglected the integrals along the string occurring in
(\ref{eq:modfder}). Again $G(x_1,x_2; y_1,y_2)$ coincides with 
$G^{\rm gi}(x_1,x_2; y_1,y_2)$ and $G(x - y)$ with $G^{\rm gi}(x - y)$
in the considered limit.
 
   Eq.'s (\ref{eq:fsqq}) and (\ref{eq:fsq}) are given in terms of purely
configurational path integrals. For the need of the following sections path
integral representations on the phase space are more convenient. These can
be obtained from the preceding ones by performing a Legendre transformation
on the exponents (see \cite{bmp} for details). At the first order in $E$, 
or equivalently at the first order in $\alpha_{\rm s} $ and $\sigma a^2$ 
($a$ being a caractheristic lenght of the problem), we have

\begin{eqnarray}
H(x_1,x_2 &;& y_1,y_2) = ({1 \over 2})^2 \int_0^{\infty} d s_1
     \int_0^{\infty} d s_2  \nonumber \\
& & \int_{y_1}^{x_1} {\cal D} z_1{\cal D} p_1\int_{y_2}^{x_2} {\cal D} z_2
    {\cal D} p_2  \exp \big \{  i 
    \sum_{j=1}^2  \int_0^{s_j} d\tau_j 
    [- \dot{z}_j p_j + {1\over 2} (p_j^2-m_j^2)] \big \} \nonumber \\
& &  \times {\cal S}_0^{s_1} {\cal S}_0^{s_2}
   \exp \big \{ i \sum_{j=1}^2 \int_0^{s_j} d \tau_j  \int_0^{\tau_j} 
   d \tau _j ^\prime 
   E( z_j -z_j^\prime ; p _j , p _j^\prime) - \nonumber\\
& & \qquad \qquad \qquad \qquad -i \int_0^{s_1} d \tau_1  \int_0^{s_2} d \tau _2 
E( z_1 -z_2 ; p_1 , p_2) \big \}
\label{eq:psqq}
\end{eqnarray}

\noindent
and

\begin{eqnarray}
H(x-y) & &= {1 \over 2} \int_0^{\infty} d s
 \int_y^x {\cal D} z {\cal D} p
 \exp \big \{  i
 \int_0^{s} d\tau [- \dot{z} p + {1\over 2} (p^2-m^2)] \big \} \nonumber \\
& & \qquad \qquad \qquad \qquad \times {\cal S}_0^s
\exp \big \{i \int_0^s  d\tau \int_0^\tau  d \tau ^\prime 
E(z - z^\prime; p , p ^\prime) \big \} \, ,
\label{eq:psq}
\end{eqnarray}

\noindent
which is justified by the fact that integration of the momenta in the 
gaussian approximation reproduces \ref{eq:fsqq} and \ref{eq:fsq} up to the 
mentioned order.


\section{The Schwinger Dyson Equation}

     From (\ref{eq:psq}) a Schwinger Dyson equation can be derived by a 
technique strictly similar to that used in \cite{bmp} to obtain a Bethe 
Salpeter equation.

     Using a trivial identity we can write with obvious notations

\begin{eqnarray}
H(x-y) &=& {1 \over 2} \int_0^{\infty} d s
   \int_y^x {\cal D} z {\cal D} p \,
   \exp \big [i \int_0^{s} d\tau K \big ] \, {\cal S}_0^s 
   \bigg \{1 + \nonumber \\
& &  + i \int_0^s d\tau \int_0^\tau d\tau^\prime 
   E(z - z^\prime; p , p ^\prime)
   \exp \big [i \int_0^\tau d \bar \tau \int_0^{\bar \tau} 
   d \bar \tau ^\prime 
   E(\bar z - \bar z^\prime;\bar p ,\bar p ^\prime) \big ] 
   \bigg \} = \nonumber\\
&=&  H_0(x-y) + {i\over 2} \int_0^{\infty} d s
 \int_y^x {\cal D} z {\cal D} p
\,\exp \big [i \int_0^{s} d\tau K \big ] \nonumber \\
& & \times \int_0^s d\tau \int_0^\tau d\tau^\prime
  J_{ab}(z - z^\prime; p , p ^\prime) \, \sigma^a  
   {\cal S}_{\tau^\prime}^\tau 
  \big \{\exp \big [i \int_{\tau ^\prime}^\tau d \bar \tau  
  \int_{\tau ^\prime}^{\bar \tau} d \bar \tau ^\prime 
  E(\bar z - \bar z^\prime;\bar p ,\bar p ^\prime) \big ] 
  \big \} \nonumber \\
& & \times \sigma^b {\cal S}_0^{\tau^\prime} \big \{ \exp 
  \big [i \int_0^{\tau ^\prime} 
  d \bar \tau \int_0^{\bar \tau} d \bar \tau ^\prime
  E(\bar z - \bar z ^\prime;\bar p ,\bar p ^\prime) \big ] \nonumber\\
& & \qquad \qquad  \times \exp \big [ i \int_{\tau^\prime}^\tau d \bar \tau 
  \int_0^{\tau ^\prime} d \bar \tau ^\prime 
  E(\bar z - \bar z^\prime;\bar p ,\bar p ^\prime) \big ] \big \} \, ,
\label{eq:prsd}
\end{eqnarray}

\noindent
where

\begin{equation}
K = -p \dot z + p^2 - m^2  \label{eq:kkk}
\end{equation}

\noindent
and

\begin{equation}
H_0(x-y)  = {1 \over 2} \int_0^{\infty} d s
 \int_y^x {\cal D} z {\cal D} p
 \exp \big [i \int_0^{s} d\tau K \big ] \ ,
\label{eq:hzero}
\end{equation}

\noindent
We have also set $a,b=0,\ \mu \nu$, with $\sigma^0=1$, and have denoted by 
$J_{ab}$ coefficients which come from $E$ and its commutation with 
${\cal S}_0^s$ and are given by (see Eq. (A8))

\begin{eqnarray}
J_{0;0} & = & E(z-z^\prime ;p,p^\prime) = \nonumber\\
   &=& 4 \pi {4\over 3} \alpha_{\rm s} 
   D_{\alpha \beta} (z-z^\prime) p^\alpha p^{\prime \beta} + \nonumber \\
   & & \qquad
   + \sigma \,\delta (z_0-z_0^\prime)\, \vert {\vec z} 
   - {\vec z}^\prime \vert
   \, \epsilon (p_0) \epsilon (p_0^\prime)  \int_0^1 d \lambda 
   \big \{ p_0^2 p_0^{\prime 2} -
   [\lambda p_0^\prime {\vec p}_{\rm T} +
   (1-\lambda) p_0 {\vec p}_{\rm T}^\prime ]^2 \big \} ^{1 \over 2}
    \nonumber \\
J_{\mu \nu ; 0} &=& - \pi {4 \over 3} \alpha_{\rm s} 
  (\delta_\mu^\alpha \partial_\nu - \delta_\nu^\alpha \partial_\mu)
  D_{\alpha \beta}(z-z^\prime) p_\beta^\prime - \nonumber \\
  & &  \qquad \qquad - \sigma \delta (z_0-z_0^\prime) \epsilon (p_0 )
  {(z_\mu-z_\mu^\prime) p_\nu -(z_\nu-z_\nu^\prime) p_\mu \over 
  \vert {\vec z}-{\vec z}^\prime \vert 
  \sqrt{p_0^2-{\vec p}_{\rm T}^2} } 
  p_0^\prime  \nonumber \\
J_{0; \rho \sigma} &=& \pi {4 \over 3} \alpha_{\rm s} 
  p^\alpha (\delta_\rho^\alpha \partial_\sigma 
  - \delta_\sigma^\alpha \partial_\rho)
  D_{\alpha \beta}(z-z^\prime) + \nonumber \\
  & & \qquad \qquad + \sigma \delta (z_0-z_0^\prime) p_0 
  {(z_\mu-z_\mu^\prime) p_\nu^\prime 
  - (z_\nu-z_\nu^\prime) p_\mu^\prime \over 
  \vert {\vec z}-{\vec z}^\prime \vert 
  \sqrt { p_0^{\prime 2}-{\vec p}_{\rm T}^{\prime 2} } } 
  \epsilon (p_0^\prime) \nonumber \\
J_{\mu \nu ; \rho \sigma} &=& - {\pi \over 4}{4\over 3} \alpha_{\rm s}
  (\delta_\mu^\alpha \partial_\nu - \delta_\nu^\alpha \partial_\mu) 
  (\delta_\rho^\alpha \partial_\sigma 
   - \delta_\sigma^\alpha \partial_\rho)
   D_{\alpha \beta}(z-z^\prime)
\label{eq:jab}
\end{eqnarray}

Up to the first order in $E$ we can replace the last exponential in 
(\ref{eq:prsd}) by 1. Then we obtain

\begin{eqnarray}
H(x-y) &=& H_0(x-y) + i \int d^4 \xi d^4 \eta d^4 \xi^\prime 
       d^4 \eta^\prime 
         H_0(x-\xi) \nonumber \\
     & & \times I_{ab}(\xi,\xi^\prime;\eta,\eta^\prime)
         \sigma^a H(\eta - \eta^\prime) \sigma^b H(\xi^\prime - y) \ ,
\label{eq:sdconf}
\end{eqnarray}

\noindent
where 

\begin{eqnarray}
 I_{ab}(\xi,\xi^\prime &;& \eta,\eta^\prime) = \nonumber \\
      & & = 4 \int {d^4 p \over (2 \pi)^4} {d^4 p^\prime \over (2 \pi)^4}
        e^{-ip (\xi - \eta)} J_{ab} ({\xi + \eta \over 2} -
        {\xi^\prime + \eta^\prime \over 2} ; p, p^\prime)
        e^{-ip^\prime( \eta^\prime - \xi^\prime)}
\label{eq:iab}
\end{eqnarray}

\noindent
and we have used the equation

\begin{equation}
\begin{array}{l}
 \int_y^x {\cal D} z {\cal D} p \,
       \exp \big [i \int_0^{s} d\tau K \big ] \dots =   
 \int d^4 \xi d^4 \eta \int {d^4 p \over (2 \pi)^4} e^{-ip(\xi-\eta)}
 \\ \nonumber  
    \qquad \qquad \qquad \qquad 
    \times \int_\xi^x {\cal D} z {\cal D} p \exp \big 
   [i \int_\tau ^s d \tau^\prime K \big ]
    \int_y^\eta {\cal D} z {\cal D} p \exp \big [ i \int_0^\tau
    d \tau^\prime K \big ]  \dots \ ,
\label{eq:fnctint}
\end{array}
\end{equation} 

\noindent
which follows immediately from the discrete form of the functional integral.

    Furthermore, by performing a Fourier trasform of Eq. (\ref{eq:sdconf}) 
we obtain

\begin{eqnarray}
\hat H(k) &=& \hat H_0(k) + \nonumber \\
& & + i \int {d^4 l \over (2 \pi)^4} \, \hat H_0(k) \,
\hat I_{ab} \Big ( k-l;{k+l \over 2},{k+l \over 2} \Big ) \,
\sigma^a \hat H(l) \, \sigma^b \hat H(k).
\label{eq:sdmom}
\end{eqnarray}

\noindent
Finally, by comparison with

\begin{equation}
\hat H(k) =\hat H_0(k) + i\hat H_0(k)\hat \Gamma (k) \hat H(k) \, ,
\label{eq:gamma}
\end{equation}

\noindent
(\ref{eq:sdmom}) gives the Schwinger Dyson equation

\begin{equation}
\hat \Gamma(k) =  \int {d^4 l \over (2 \pi)^4}  \,
\hat I_{ab} \Big ( k-l;{k+l \over 2},{k+l \over 2} \Big ) 
\sigma^a \hat H(l) \, \sigma^b \ .
\label{eq:sdeq}
\end{equation}

In an explicit form the kernels $\hat I_{ab}$ are given by

\begin{eqnarray}
\hat I_{0;0} (Q; p, p^\prime) & = & 4 \int d^4 \zeta \, e^{iQ \zeta} 
   E(\zeta ;p,p^\prime) = 
   16 \pi {4 \over 3} \alpha_{\rm s} p^\alpha p^{\prime \beta}  
  \hat D_{\alpha \beta} (Q)  + \nonumber \\ 
 &+& 4 \sigma  \int d^3 {\vec \zeta} e^{-i{\vec Q}\cdot {\vec \zeta}} 
    \vert {\vec \zeta} \vert \epsilon (p_0) \epsilon ( p_0^\prime )
   \int_0^1 d \lambda \{ p_0^2 p_0^{\prime 2} -
   [\lambda p_0^\prime {\vec p}_{\rm T} +
   (1-\lambda) p_0 {\vec p}_{\rm T}^\prime ]^2 \} ^{1 \over 2} \nonumber \\
\hat I_{\mu \nu ; 0}(Q;p,p^\prime) &=& 4\pi i {4 \over 3} \alpha_{\rm s} 
   (\delta_\mu^\alpha Q_\nu - \delta_\nu^\alpha Q_\mu) p_\beta^\prime
   \hat D_{\alpha \beta}(Q)  - \nonumber \\
   &-& \sigma  \int d^3 {\vec \zeta} \, e^{-i \vec Q \cdot \vec \zeta} 
   \epsilon (p_0)
   {\zeta_\mu p_\nu -\zeta_\nu p_\mu \over 
   \vert {\vec \zeta} \vert \sqrt{p_0^2-{\vec p}_{\rm T}^2}} 
   p_0^\prime  \nonumber \\
\hat I_{0; \rho \sigma}(Q;p,p^\prime) &=& 
   -4 \pi i{4 \over 3} \alpha_{\rm s} 
   p^\alpha (\delta_\rho^\beta Q_\sigma - \delta_\sigma^\beta Q_\rho)
   \hat D_{\alpha \beta}(Q) + \nonumber \\
   &+& \sigma  \int d^3 {\vec \zeta} \, e^{-i{\vec Q} 
  \cdot {\vec \zeta}} p_0 
  {\zeta_\rho p_\sigma^\prime - \zeta_\sigma p_\rho^\prime \over 
  \vert {\vec \zeta} \vert \sqrt{p_0^{\prime 2}
   -{\vec p}_{\rm T}^{\prime 2}} } 
  \epsilon (p_0^\prime)  \nonumber \\
\hat I_{\mu \nu ; \rho \sigma}(Q;p,p^\prime) &=&  
   \pi {4\over 3} \alpha_{\rm s}
  (\delta_\mu^\alpha Q_\nu - \delta_\nu^\alpha Q_\mu) 
  (\delta_\rho^\alpha Q_\sigma - \delta_\sigma^\alpha Q_\rho)
  \hat D_{\alpha \beta}(Q) 
\label{eq:imom}
\end{eqnarray}

\noindent
where in the second and in the third equation $\zeta_0 = 0$ has to be 
understood. 

    Notice that, were \ref{eq:sdeq} covariant, i.e. were the quantities 
$ \hat I_{ab}$ tensors, we could write

\begin{equation}
i \hat H^{-1}(k) = a(k^2) +  \not k b(k^2) \, .
\label{eq:hreca}
\end{equation}

\noindent
However, due to the equal time straight line approximation, that is 
not the case; we must assume (\ref{eq:imom}) to hold in the meson center
of mass frame and set

\begin{equation}
i \hat H^{-1}(k) = \sum_{r=0}^3 \omega_r(k) h_r(k) \, ,
\label{eq:hrecb}
\end{equation}

\noindent
with

\begin{equation}
\omega_0 = 1\, , \ \ \ \ \omega_1 = \gamma^0 \, , \ \ \ \
\omega_2 = - \vec \gamma \cdot \hat k \, , \ \ \ \
\omega_3 = \gamma^0 \vec \gamma \cdot \hat k  \, , 
\label{eq:omega}
\end{equation}

\noindent
$\hat k = {1 \over |\vec k|} \vec k $ and $h_0(k), 
\dots h_3(k)$ functions of $k_0$ and $|\vec k|$ separately. 
Obviously (\ref{eq:hrecb}) reduces to (\ref{eq:hreca}) 
for $ h_0 = a(k^2), \ \
h_1 = k_0 \, b(k^2) , \ \ h_2 = |\vec k | \, b(k^2), \ \ h_3 = 0 $.

   From (\ref{eq:gamma}) and (\ref{eq:hrecb}) it follows

\begin{equation}
\hat \Gamma (k) = i \hat H^{-1}(k)-i \hat H_0^{-1}(k) 
   = h_0(k)-(k^2-m^2)
    +\sum_{r=1}^3 \omega_r (k) h_r(k) 
\label{eq:gammac}
\end{equation}

\noindent
and

\begin{equation}
\hat H(k) = i { h_0(k) - \sum_{r=1}^3 \omega_r (k) h_r(k)  
     \over h_0^2(k) - h_1^2(k) + h_2^2(k) -h_3^2(k)} \, .
\label{eq:homega}
\end{equation}

   Replacing such equations in (\ref{eq:sdeq}) and taking advantage of

\begin{equation}
{1 \over 4} {\rm Tr} (\omega_r^+ \omega_s) = \delta_{rs} \, ,
\label{eq:ortoomega}
\end {equation}

\noindent
we finally obtain

\begin{equation}
h_r (k) = \delta_{r0} (k^2-m^2) - i \sum_{s=0}^3 \int {d^4 l \over (2 \pi)^4} 
    {R_{rs}(k,l) \, h_s(l) \over h_0^2(l) - h_1^2(l) + h_2^2(l) -h_3^2(l)} \, ,
\label{eq:sdexpl}
\end{equation}

\noindent
with

\begin{equation}
R_{rs}(k,l) = \mp {1 \over 4} \hat I_{ab} (k-l;{k+l \over 2}, {k+l \over 2})
      {\rm Tr} [ \omega_r^+(k) \sigma^a  \omega_s(k) \sigma^b ] \, ,
\label{eq:kernexpl}
\end{equation}

\noindent
where the sign - applies to the $s=0$ case, the sign + to all the other 
cases. Notice that from well known properties of Dirac's matrices it
follows 
immediately $ R_{01} = R_{10} = R_{02} = R_{20} = R_{13} = R_{31} =
R_{23} = R_{32} = 0 $, while from

\begin{equation}
\hat I_{\mu \nu ; 0}(Q; q,q) = - \hat I_{0; \mu \nu}(Q; q,q) \ \
         {\rm and} \ \
        \hat I_{\mu \nu ; \rho \sigma}(Q; q,q)  =
         \hat I_{\rho \sigma; \mu \nu}(Q; q,q)
\label{eq:isimm}
\end{equation}

\noindent
it follows also $ R_{03} = R_{30} = 0 $. Actually only $ R_{00}, \  R_{11},
\ R_{12}, \ R_{21}, \ R_{22}, \  R_{33} $ are different from zero.
Expressions for these quantities in terms of the $ \hat I_{ab} $ are given 
in Appendix B. Here, as an example, we want only to report the explicit 
expression

\begin{equation}
\begin{array}{ll}
R_{00}(k,l)  & = - 4 \pi {4 \over 3} \alpha_{\rm s}[4{p^2 l^2 -
    (pl)^2 \over (p-l)^4}+{3 \over 4}] - \nonumber \\
     & \qquad - \sigma  \int d^3 \vec \zeta e^{-i(\vec k - \vec l) \cdot \vec \zeta} 
    \vert {\vec \zeta} \vert  \vert( k_0 + l_0)\vert \sqrt { (k_0 +l_0)^2
    - (\vec k_{\rm T} + \vec l_{\rm T} )^2 } 
\end{array} \, . \label{eq:krsing}
\end{equation}

\noindent
The confining term in this equation contains an infrared singularity 
essentially of the type

\begin{equation}
{\partial^2 \over \partial \epsilon^2} \, {1 \over ({\vec k} - {\vec l})^2 + 
        \epsilon^2} \, ,
\label{eq:infsing}
\end{equation}

\noindent
with $\epsilon \to 0 $. Similar singularities appear in the other diagonal 
kernels $R_{rr} $. 

    Expressions like (\ref{eq:infsing}) correspond to well 
defined distributions and are harmless if they appear in integrals
multiplied by regular functions. However, they can give troubles and 
generate new singularities when occurring in connection with 
discontinuous functions. As it appears from (\ref{eq:krsing}) and 
(\ref{eq:imom}), $ R_{rs} $ have also a bad ultraviolet behaviour that
should be handled by renormalization. Renormalization would affect the
perturbative parts of $ R_{rs} $, but its explicit consideration is not
important for our present purpose. So we shall simply 
suppose to regulate (\ref{eq:sdexpl}) by a cut-off $\Lambda$.

     Finally going back to the first order propagator (\ref{eq:qmdfeqg})
we can also set

\begin{equation}
iG^{-1}(k) = \sum_{r=0}^3 \omega_r(k) g_r(k) \, ,
\label{eq:grec}
\end{equation}

\noindent
and then we have

\begin{equation}
\left \{
\begin{array}{l}
h_0 = m g_0 + k_0 g_1 + |\vec k| g_2  \nonumber \\
h_1 = m g_1 + k_0 g_0 - |\vec k| g_3  \nonumber \\
h_2 = m g_2 - |\vec k| g_0 + k^0 g_3  \nonumber \\
h_3 = m g_3 + |\vec k| g_1 + k_0 g_2 
\end{array} \right.   
\label{eq:gexpl}
\end{equation}

From (\ref{eq:grec}) it is apparent that, for zero quark masses, chiral 
symmetry breaking corresponds to $ g_0(k) \not= 0 $ and/or $  g_3(k)
\not= 0 $ and consequently from (\ref{eq:gexpl}) to $h_1(k)$ and 
$h_2(k)$ not simultaneously vanishing.


\section{The Bethe-Salpeter equation}
\indent
    The Bethe-Salpeter equation for the quantity $H(x_1,x_2;y_1,y_2)$ 
is derived from (\ref{eq:fsqq}) by applying the same identity used in 
(\ref{eq:prsd}) to the interaction, rather than to the selfenergy term,
as done in the case of the Schwinger-Dyson equation. In terms of the 
quantities defined by Eq.'s (\ref{eq:iab}) and (\ref{eq:jab}) it reads

\begin{eqnarray}
H(x_1,x_2;y_1,y_2) &=& H_1(x_1-y_1) \, H_2(x_2-y_2) - \nonumber \\
    & & - i \int d^4 \xi_1 d^4 \xi_2 d^4 \eta_1 d^4 \eta_2 
    H_1(x_1-\xi_1) \, H_2(x_2-\xi_2) \nonumber \\
    & & \qquad \qquad \times I_{ab}(\xi_1,\xi_2;\eta_1,\eta_2)\, 
    \sigma_1^a \, 
    \sigma_2^b \, H(\eta_1,\eta_2;y_1,y_2) \, ,
\label{eq:bseq}
\end{eqnarray}

\noindent
where $H_1$ and $H_2$ denote the quark and the antiquark propagators 
as defined by (\ref{eq:fsq}).

   The corresponding homogenuous equation in the momentum 
representation is

\begin{eqnarray}
\Phi_ P(k) &=& -i \int {d^4u \over (2 \pi)^4} \hat H_1 
   \big ({1 \over 2} P 
   + k \big ) \hat H_2 \big ({1 \over 2} P - k \big ) \nonumber \\
   & & \qquad \qquad \qquad \hat I_{ab} \big (k-u, {1 \over 2}P
   +{k+u \over 2}, 
   {1 \over 2}P-{k+u \over 2} \big )
    \sigma_1^a \sigma_2^b \Phi_P (u) \, ,
\label{eq:bshoma}
\end{eqnarray}

\noindent
the center of mass frame being understood, i.e. $P=(m_B, {\vec 0})$.

   The wave function $\Phi_P (k)$ can be also reinterpreted as a matrix
(in which the column index refers to the quark and the row one to the 
antiquark respectively) and the pedices 1 and 2 in the spin operators 
suppressed. Then setting

\begin{equation}
\Phi_P (k) = \hat H_1 \big ({1 \over 2} P + k \big ) \, \Gamma_P^M(k) \,
       \hat H_2 \big (-{1 \over 2} P + k \big ) C
\label{eq:wfnct}
\end{equation}

\noindent
and using $C \hat H^{\rm T}(k) C^{-1} = \hat H^{\rm T}(-k)$ we can rewrite 
(\ref{eq:bshoma}) as

\begin{eqnarray}
\Gamma_P^M(k) &=& -i \int{d^4u \over (2 \pi)^4}    
   \hat I_{ab} \big (k-u, {1 \over 2}P+{k+u \over 2}, {1 \over 2}P-
    {k+u \over 2} \big ) \nonumber \\ 
   & & \qquad \qquad \qquad \qquad \sigma^a \, 
    \hat H_1({1 \over 2} P + u) \, 
   \Gamma_P^M(u) \, \hat H_2 \big (-{1 \over 2} P + u \big ) \, 
    \bar \sigma^b
\label{eq:bshomb}
\end{eqnarray}

\noindent
with

\begin{equation}
\bar \sigma^0 = \sigma^0 \ , \quad \quad \quad \quad \bar \sigma^{\mu \nu}=
   C (\sigma^{\mu \nu})^{\rm T} C^{-1} = - \sigma^{\mu \nu} \ ,
\label{eq:brsgm}
\end{equation}

\noindent
$C$ being the charge conjugation matrix. This is the form of the BS-equation 
more convenient for our present purposes. 

   More specifically we need to consider the case $P=0$, corresponding to a 
zero mass bound state. If we take into 
account the property (cf. (\ref{eq:imom}))

\begin{equation}
\hat I_{ab}(Q;p,p^\prime) \bar \sigma^b =
       - \hat I_{ab}(Q;p,-p^\prime)  \sigma^b \, ,
\label{eq:iprprt}
\end{equation}

\noindent
and assume $m_1=m_2=m$, we have

\begin{equation}
\Gamma_0^M(k) = i \int{d^4u \over (2 \pi)^4}    
   \hat I_{ab} \big (k-u, {k+u \over 2}, {k+u \over 2} \big )
   \sigma^a \hat H(u)\Gamma_0^M(u) \hat H(u)
   \sigma^b \ .
\label{eq:bshomc}
\end{equation}

\noindent
Notice that a zero bound state exists, if (\ref{eq:bshomc}) has a non trivial 
solution.


\section{Chiral symmetry breaking an Goldstone theorem}

\indent
  To discuss Eq.'s (\ref{eq:sdeq}) or(\ref{eq:sdexpl}) and 
(\ref{eq:bshomc}) it is convenient to 
rewrite them in euclidean form. This can be obtained by making the 
substitutions $p_0 \to ip_4$ in all momentum variables 
and  by setting $\gamma_0 = i \gamma_4 $. To solve the ambiguities 
related to the occurrence sign funtions and the square roots  
in (\ref{eq:imom}) one should go back to the 
eucleadean counterparts of the original equations and particularly of 
(\ref{eq:psqq}) and (\ref{eq:psq}). Alternatively one can extract from 
the square roots factors as $|p_0|$ and then set $\epsilon (p_0) 
|p_0| = p_0$ before performing the substitution. It also convenient to
redefine $h_1$ as $ih_1$, $h_3$ as $ih_3$ and also $R_{12}$ as 
$iR_{12}$ and $R_{21}$ as $-iR_{21}$. 
The resulting expression for $R_{00}$ is, e.g.,

\begin{equation}
\begin{array}{ll}
R_{00}(k,l)  & = - 4 \pi {4 \over 3} \alpha_{\rm s}[4{p^2 l^2 -
    (pl)^2 \over (p-l)^4}+{3 \over 4}] + \nonumber \\
 & \qquad + \sigma  \int d^3 \vec \zeta e^{-i(\vec k - \vec l) 
    \cdot \vec 
   \zeta} \vert {\vec \zeta} \vert  \vert( k_4 + l_4)\vert 
    \sqrt { (k_4 +l_4)^2
    + (\vec k_{\rm T} + \vec l_{\rm T} )^2 } 
\end{array} \, . \label{eq:k00eucl}
\end{equation}
 
\noindent
while expressions for the confinement parts of the other $R_{rs}$ 
are reported in appendix C.  

    Then, making explicit the terms that are different from zero, 
(\ref{eq:sdexpl}) can be rewritten

\begin{equation}
\left \{
\begin{array}{l}
h_0(k) = -k^2-m^2 + \int {d^4 l \over (2 \pi)^4 }
    {R_{00}(k,l) \, h_0(l)  \over h_0^2(l) 
    + h_1^2(l) + h_2^2(l) +h_3^2(l)}   \nonumber \\
h_1(k) = \int {d^4 l \over (2 \pi)^4} 
    {R_{11}(k,l) \, h_1(l) + R_{12}(k,l) \, h_2(l) \over h_0^2(l) 
    + h_1^2(l) + h_2^2(l) + h_3^2(l)} 
  \nonumber \\
h_2(k) = \int {d^4 l \over (2 \pi)^4} 
    { R_{21}(k,l) \, h_1(l) + R_{22}(k,l) \, h_2(l) \over h_0^2(l) 
    + h_1^2(l) + h_2^2(l) + h_3^2(l)}    \nonumber \\
h_3(k) =  \int {d^4 l \over (2 \pi)^4 }
    {R_{33}(k,l) \, h_3(l)  \over h_0^2(l) 
    + h_1^2(l) + h_2^2(l) + h_3^2(l)}  
\end{array} \right.   
\label{eq:sdeucl}
\end{equation}

Similarly for the euclidean counterpart of (\ref{eq:bshomc}) we have

\begin{eqnarray}
\Gamma_0^M  (k) &=&  \int{d^4u \over (2 \pi)^4}    
   \hat I_{ab} \big (k-u, {k+u \over 2}, {k+u \over 2} \big ) 
     \sigma^a { h_0(u) + \gamma_4 h_1(u) + \vec \gamma
        \cdot \hat u h_2(u) + \gamma_4 \vec \gamma \cdot \hat u 
          \, h_3(u) \over h_0^2(u) + h_1^2(u) + h_2^2(u) + h_3^2(u)} 
        \nonumber \\
  & & \qquad \qquad \qquad \qquad \Gamma_0^M(u) 
         {h_0(u) + \gamma_4 h_1(u) +  \vec \gamma
         \cdot \hat u h_2(u) + \gamma_4 \vec \gamma \cdot \hat u 
         \, h_3(u) \over h_0^2(u) + h_1^2(u) + h_2^2(u) + h_3^2(u)}
 \sigma^b \ .
\label{eq:bseucl}
\end{eqnarray}

    Notice that {\it a priori} Eq. (\ref{eq:sdeucl}) could be expected to 
admit various types of solutions corresponding to different sets of $h_s$
identically vanishing

\begin{equation}
\begin{array}{l}
{\rm A}1) \qquad \qquad h_1=h_2=h_3=0\, , \nonumber \\
{\rm A}2) \qquad \qquad h_1, \, h_2 \not= 0 , \ \ \ h_3=0\, , \nonumber \\
{\rm A}3) \qquad \qquad h_1=h_2=0, \ \ \ h_3 \not= 0 \, , \nonumber \\
{\rm A}4) \qquad \qquad h_1, \,h_2, \,h_3 \not= 0\, .
\end{array}
\end{equation}

To gain more insight on the nature of such solutions let us assume 
that we can solve (\ref{eq:sdeucl}) by iteration starting by an initial 
ansatz of the type 

\begin{equation}
\qquad \qquad \qquad  h_0^{(0)}=-k^2-m^2\, ,   \qquad \qquad  h_r^{(0)}= 
     \mu_r^2 f_r \big ({k \over \bar \mu} 
     \big)\, , \qquad  \qquad \qquad  (r=1,2,3) 
\label{eq:startingstep}
\end{equation}

\noindent
$\mu_1,\mu_2,\mu_3$ and $\bar \mu$ being constants with the dimensions 
of a mass and $f_r$ arbitrary functions chosen in such
a way that at least the first iteration is meaningfull. E. g. after the 
first iteration we have for $h_0$

\begin{equation}
h_0^{(1)}(k) = - k^2-m^2 - \int {d^4 l \over  (2 \pi)^4} 
    {R_{00}(k,l) \, (l^2+m^2)  \over  (l^2+m^2)^2 
    + [h_1^{(0)}(l)]^2 + [h_2^{(0)}(l)]^2 + [h_3^{(0)}(l)]^2}  \, .
\label{eq:itera}
\end{equation}

Solutions of the types A1-A4) can be obtained from (\ref{eq:startingstep}) 
simply by taking $\mu_1=\mu_2=\mu_3=0$, $\mu_1, \, \mu_2 \not= 0, \ \mu_3=0$ 
etc. respectively. Notice that $\mu_1, \dots \bar \mu$, when different from zero
must be expressed in terms of the masses existing in the theory, $m$, 
$\sqrt{\sigma}$ and the cut-off $\Lambda$ or the renormalization scale.

   Let us consider separately the various cases.

   {\it  Case} A1).  In this case we expect no actual solution smooth 
in $m=0$. In fact, setting $m=0$ and $h_1=h_2=h_3=0$ Eq. 
(\ref{eq:itera}) becomes

\begin{equation}
h_0^{(1)}(k) = - k_4 ^2 -\vec k^2- \int {d^4 l \over (2 \pi)^4 }
    {R_{00}(k,l)  \over  l_4 ^2 + \vec l^2}  \, .
\label{eq:iterb}
\end{equation}

\noindent 
Having in mind Eq. (\ref{eq:krsing}) we realize that integration over 
$l_4$ in such an equation produces a ${1 \over |\vec l|}$ singularity
in the tridimensional integral which conspires with (\ref{eq:infsing})
for $\vec k \to 0$ bearing a ${1 \over \vec k ^2}$ singularity in 
$h_0^{(1)}(k)$. 
Similar circustances occurre at various steps in the iteration rendering
the entire process instable and apparently preventing convergence. 

   {\it Case} A2). One can immmediately check that, if $h_3$ vanishes,
setting

\begin{equation}
\Gamma_0^M(k) = [\gamma_4 h_1(k) + \vec \gamma \cdot \hat k
    h_2(k)] \gamma_5 \, ,
\label{eq:goldstone}
\end{equation}

\noindent
(\ref{eq:bseucl}) is made identical to the subsystem formed by the 
second and the third Eq. (\ref{eq:sdeucl}) and so non trivially 
verified. Therefore, a zero mass pseudoscalar meson exists, 
independently of the quark mass value and of the occurrence of a 
specific symmetry. This class has to be rejected. 

{\it Case} A3). This time a non trivial solution of
(\ref{eq:bseucl}) can be obtained by setting 

\begin{equation}
\Gamma_0^M(k) = \gamma_4 h_3(k) \, .
\label{eq:nogold}
\end{equation}

\noindent
This would correspond to a zero mass scalar meson. Also this class
must be considered spurious.

    We are left with case A4 in which all the quantities $h_1$, $h_2$ 
and $h_3$ are different from zero. Inside this class we can still
distinguish various, possibly single, solutions corresponding to 
different behaviours of the $h_s$ in the limit $m \to 0$. Schematically
we can think of generating such different solutions by setting in 
({\ref{eq:startingstep})

\begin{equation}
\begin{array}{l}
{\rm B}1) \qquad \qquad \mu_1=\mu_2=\mu_3=m \, , \nonumber \\
{\rm B}2) \qquad \qquad \mu_1=\mu_2=\sqrt{\sigma}\, , \ \ \ \mu_3=m \, , 
                                  \nonumber \\
{\rm B}3) \qquad \qquad \mu_1=\mu_2=m, \ \ \ \mu_3=\sqrt{\sigma} \, , 
                                  \nonumber \\
{\rm B}4) \qquad \qquad \mu_1=\mu_2=\mu_3=\sqrt{\sigma}\, ,
\end{array}
\end{equation}

   Notice that B1, B2, B3 correspond to solutions that are reduced
respectively to the types A1, A2, A3 for $m \to 0$; in fact the factors 
$m$ in front of the various $f_r(k)$ are reproduced at every step of 
the iteration. Notice also that for $m \to 0$ the theory acquires chiral
symmetry. As we mentioned the symmetry is broken if in the same limit 
$h_1$ and $h_2$ do not simultaneously vanish. So we have a broken chiral
symmetry in cases B2 and B4, actual symmetry in the other two cases.

   Let us consider again the various cases.

   Ansatzes B1 and B3 reproduce similar situations as A1 and A3 for 
$m \to 0$ and can be excluded for the same reasons. Ansatz B4 
contradicts Goldstone's theorem; in fact for $m \to 0$, $h_3$ 
keeps different from zero in this case and the expression 
(\ref{eq:goldstone}) would not be a solution of (\ref{eq:bseucl}).
On the contrary ansatz B2 seems to satisfy all requirements. For 
$m \to 0$ (\ref{eq:goldstone}) becomes a non trivial solution of 
(\ref{eq:bseucl}) and and a zero mass pseudo scalar bound state occurs.

   In conclusion we are left with solution B2 alone. This seems to be
the only one which is physically sensible and mathematically consistent.
For light $u$ and $d$ quarks it would correspond to a breaking of the 
approximate chiral 
symmetry and would correctly provide a $\pi$ meson with a small mass.

\section{Conclusions}

\indent
    Using the formalism developped in \cite{bmp}, the minimal area law
and the straight line approximation in a modified form, we have 
obtained the Schwinger-Dyson equation (\ref{eq:sdeq}) for the quark
propagator occurring in the Bethe-Salpeter equation (\ref{eq:bseq}) or
(\ref{eq:bshomb}). We have 
explicitly rewritten the SD equation as a system of non linear integral 
equations involving four scalar functions. However, we found that such 
system has not a unique solution and had to resort to additional 
criteria to select the correct one. 

    After a first selection, assuming an iterative resolution 
procedure to converge, we have been brought to consider various possible
solutions corresponding to different starting ansatzes. One
of these (solution B2) produces chiral symmetry breaking in the 
quark mass vanishing limit and a zero mass pseudoscalar meson. This
corresponds to what we believe to be the real situation. All the other
solutions are mathematically inconsitent, or in contradiction with 
general theorems, or, simply, have unexpected features which we do not 
believe should be present in real QCD. If we accept B2 to be the correct
solution, then, we achieve a very consistent formalism for the treatment
of the quark-antiquark 
bound states in QCD, both for light and for heavy quarks.

   We stress once more that the occurrence in the kernels of the infrared
singularity (\ref{eq:infsing}) is what seems to prevent solutions A1 
or B1 form existing. Therefore a strict connection appears to exist 
between confinement as expressed 
by the area law and chiral symmetry breaking.

     Notice, also, that the possibility of a solution with the properties 
we have found depends strictly on the assumption of (\ref{eq:nstrai}) 
in place of (\ref{eq:lstrai}). Had we used straight line approximation
in the original form, only the perturbative parts of (\ref{eq:imom}) 
would have occurred in (\ref{eq:sdeq}). Then the kernels in 
(\ref{eq:bshomc}) would have been different from the kernels in 
(\ref{eq:imom}) and (\ref{eq:goldstone})
would not have been a solution of (\ref{eq:bshomc}). The correct solution 
of (\ref{eq:sdeucl}) would have been of the type A1 and chiral symmetry 
breaking and the zero mass pseudo scalar meson would not have occurred. 

     Let us make a final comment.

     As we mentioned the confinement parts of the kernels $\hat 
I_{ab}$, as given by (\ref{eq:imom}), are not tensors. As we mentioned, 
this circustance is a consequence of the equal time straight line 
approximation, which is obviously frame dependent. We have assumed 
the priviledged frame to be the center of mass frame of the meson. 
But then $\hat I_{ab}$ can be rewritten as tensors, if we introduce 
explicitly the total four-momentum of the meson $P$. For this aim 
it is sufficient to set

\begin{equation}
Q_{\|}^\rho = {Q \cdot P \over P^2} P^\rho , \ \ \ \ \ \ Q_{\bot}^\rho 
       = Q^\rho - Q_{\|}^\rho,
\end{equation}
give analogous definitions for $p$ and $p^\prime$ and make the identifications
 
\begin{equation}
Q_0^2 \to Q_{\|} ^2, \ \ \ \ \ \ {\vec Q}^2 \to - Q_{\bot}^2 \ \ \ \ 
 {\rm etc.} \end{equation}

\noindent
and similar ones.

    In the above perspective the quark propagator $\hat H(k)$ should 
be rewritten as $\hat H(P,k)$ and the Eq. (\ref{eq:hrecb}) as

\begin{equation}
i \hat H^{-1}(P,k) = \sum_{r=0}^3 \omega_r^\prime (P,k) 
   h_r^\prime (P,k) \, ,
\end{equation}

\noindent
with

\begin{equation}
\omega_0^\prime = 1,\ \ \ \omega_1^\prime 
     = {1 \over \sqrt{P^2}} \not P,
     \ \ \ \omega_2^\prime = {1 \over \sqrt{k^2}} \not k, \ \ \
     \omega_3^\prime = {-i \over 2 \sqrt{P^2k^2}}[\not P, \not k]
\end{equation}

\noindent
and (\ref{eq:sdeucl}) correspondingly modified.

    Since the quark propagator has been originally defined in an 
independent way by (\ref{eq:qeqh}), at first sight it can be surprising
that now it is made dependent on the meson momentum. Notice, however, 
that it aquires again $P$ dependence as a consequence of the straight 
line approximation. After that 
it becomes the specific propagator to be used in that BS equation.

    Notice also that the need to choose solution B2 (with $h_3\not= 0$
for $m \not= 0$) is a consequence of the above dependence. Had $\hat 
I_{ab}$'s been tensors in $k$ and $l$ alone (as in the abelian case),
we should have had $ h_3= 0$ (see Eq. (\ref{eq:hreca}))
and A1 would have been the only possible sensible solution. Therefore 
a $P$ dependence of the quark propagator seems to be essential for a 
consistent chiral symmetry breaking and not an artifact of the minimal
area law model or of the straight line approximation.

\section*{Acknowledgments}
\indent
    I would like to thank C, D. Roberts, D. Gromes and N. Brambilla for 
enlightning conversations.


\appendix

\section{Functional derivatives}
\indent
     Taking into account appendix C of \cite{bmp}, 
and specifically Eq. (C10),
we can write for the perturbative part

\begin{eqnarray}
{\delta \over \delta S^{\mu \nu}(z(\tau))} && \int_a^b d \bar \tau 
       \int_{a^\prime}
       ^{b^\prime} d \bar \tau ^\prime E_{\rm pert}
        (\bar z - \bar z^\prime, 
        \dot{\bar z}, \dot{\bar z} ^\prime ) =  \nonumber \\
= && 4\pi {4 \over 3} \alpha_{\rm s}{\delta \over 
        \delta S^{\mu \nu}(z(\tau))}
        \int_a^b d \bar \tau  \int_{a^\prime} ^{b^\prime} 
d \bar \tau ^\prime 
        D_{\alpha \beta}(\bar z - \bar z ^\prime) \dot{\bar z}^\alpha 
        \dot{\bar z}^{\prime \beta} = \nonumber \\
= && 4\pi {4 \over 3} \alpha_{\rm s} \bigg \{ \chi_{(a,b)}(\tau)
        \int_{a^\prime} ^{b^\prime} d \bar \tau ^\prime 
       (\delta_\mu^\alpha \partial_\nu - \delta_\nu^\alpha \partial_\mu ) 
       D_{\alpha \beta}(\bar z - \bar z ^\prime)  
        \dot{\bar z}^{\prime \beta} - \nonumber \\
&& -  \chi_{(a^\prime,b^\prime)}(\tau)
        \int_a ^b d \bar \tau (\delta_\mu^\beta \partial_\nu -
        \delta_\nu^\beta \partial_\mu ) D_{\alpha \beta}(\bar z 
        - \bar z ^\prime) \dot{\bar z}^ \alpha \bigg \} \, ,
\end{eqnarray}

\noindent
where  $\chi_{(a,b)}(\tau)$ and  $\chi_{(a^\prime,b^\prime)}(\tau)$ are the
characteristic functions of the intervals $(a,b)$ and $(a^\prime, b^\prime)$
($(a,b)$ and $(a^\prime, b^\prime)$ contained in (0,s)) and
the derivatives are intended to act on the entire argument of 
$D_{\alpha \beta}$. Due to the arbitrarity of the above intervals, we can also
write formally

\begin{equation}
\begin{array}{rl}
{\delta \over \delta S^{\mu \nu}(z(\tau))} & 
        E_{\rm pert}(\bar z - \bar z^\prime, 
        \dot{\bar z}, \dot{\bar z} ^\prime )  =  \nonumber \\
 = & 4\pi {4 \over 3} \alpha_{\rm s} \big [\delta (\tau - \bar \tau)
       (\delta_\mu^\alpha \partial_\nu 
       - \delta_\nu^\alpha \partial_\mu ) 
       D_{\alpha \beta}(\bar z - \bar z ^\prime)  
        \dot{\bar z}^{\prime \beta} - \nonumber \\
 & \qquad \qquad - \delta( \tau - \bar \tau ^\prime) 
      (\delta_\mu^\beta \partial_\nu -
        \delta_\nu^\beta \partial_\mu) D_{\alpha \beta}(\bar z 
      - \bar z ^\prime) \dot{\bar z}^ \alpha \big ] \, .
\end{array}
\end{equation}

     For the confining part for consistency we must define

\begin{equation}
{\delta \over \delta S^{\mu \nu}(z(\tau))} \int_a^b d \bar \tau 
       \int_{a^\prime}
       ^{b^\prime} d \bar \tau ^\prime E_{\rm conf} 
       (\bar z - \bar z^\prime, 
        \dot{\bar z}, \dot{\bar z} ^\prime )
\end{equation}

\noindent
as the the straight line approximation of $ {\delta 
 \over \delta S^{\mu \nu}} S_{a a^\prime}^{b b^\prime}$, the quantity
$ S_{a a^\prime}^{b b^\prime}$ denoting the area of the strip 
of the minimal area enclosed by $\Gamma_q$ delimited by the two 
geodetics connecting $z(a)$ to $z(a^\prime)$ and  $z(b)$ to  
$z(b^\prime)$ respectively. Then, having again in mind Eq. (C9)
of \cite{bmp}, we have

\begin{equation}
\begin{array}{ll}
{\delta \over \delta S^{\mu \nu}(z(\tau))} & \int_a^b d \bar \tau 
       \int_{a^\prime}
       ^{b^\prime} d \bar \tau ^\prime E_{\rm conf}
       (\bar z - \bar z^\prime, 
        \dot{\bar z}, \dot{\bar z} ^\prime) = \nonumber \\
 & = \sigma \chi_{(a,b)}(\tau)\chi_{(a^\prime,b^\prime)}(\tau ^\prime) 
       \epsilon(\dot z_0) 
       \epsilon(\dot z_0^\prime) \bigg ( {(z^\mu
       - z^{\prime \mu})\dot z ^\nu -
      (z^\nu- z^{\prime \nu})\dot z ^\mu \over  
       [-\dot z ^2(z-z^\prime)^2 +
       (\dot z (z-z^\prime))^2]^{1 \over 2}}- \nonumber \\
& \qquad \qquad \qquad - {(z^\mu- z^{\prime \mu})\dot z^ {\prime \nu} -
      (z^\nu- z^{\prime \nu})\dot z ^{\prime \mu} \over 
      [-\dot z ^{\prime 2}
       (z-z^\prime)^2 +
       (\dot z^\prime (z-z^\prime))^2]^{1 \over 2}} \bigg ) 
       \bigg \vert _{z_0(\tau^\prime) = z_0(\tau)} = \nonumber \\
&= \sigma \bigg \{ \chi_{(a,b)}(\tau) \int _{a^\prime}^{b^\prime} 
      d \bar\tau^\prime  \delta(z_0 - \bar z_0^\prime)
       |\dot {\bar z}_0^\prime| \epsilon({\dot z}_0)
      \epsilon(\dot {\bar z}_0^\prime)
       {(z^\mu- \bar z^{\prime \mu})\dot z ^\nu -
      (z^\nu - \bar z^{\prime \nu})\dot z ^\mu \over 
       [\dot z ^2(\vec z - \vec{\bar z}^\prime)^2 +
       (\dot z (\vec z - \vec{\bar z}^\prime))^2]^{1 \over 2}} 
        - \nonumber \\
& \qquad \qquad \qquad - \chi_{(a^\prime,b^\prime)}(\tau) 
       \int _a^b d \bar\tau 
      \delta(\bar z_0 - z_0)
       |\dot {\bar z}_0| \epsilon(\dot{\bar z_0})\epsilon(\dot z_0)
       {(\bar z ^\mu - z^ \mu)\dot z ^\nu -
       (\bar z^\nu - z^ \nu)\dot z ^\mu \over [\dot z ^2(\vec{\bar z}
        - \vec z)^2
       + (\dot z (\vec{\bar z} - \vec z))^2]^{1 \over 2}} \bigg \}
\end{array}
\end{equation}

\noindent
or

\begin{equation}
\begin{array} {rl}
{\delta \over \delta S^{\mu \nu}(z(\tau))} & 
       E_{\rm conf}(\bar z - \bar z^\prime, 
        \dot{\bar z}, \dot{\bar z} ^\prime) = \nonumber \\
= & \sigma \delta (\tau - \bar \tau) 
      \delta(z_0 - \bar z_0^\prime)
       \epsilon(\dot z_0) \dot{\bar z} _0^\prime
       {(\bar z^\mu - \bar z^{\prime \mu}) \dot{\bar z} ^\nu -
      (\bar z ^\nu - \bar z^{\prime \nu})\dot {\bar z} ^\mu 
      \over | \bar {\vec z} - \bar{\vec z}^\prime| 
      \sqrt {\dot {\bar z_0}^2 - 
      \dot {\bar {\vec z}}_{\rm T}^2}} - \nonumber \\ 
& \qquad \qquad - \sigma \delta (\tau - \bar \tau ^\prime) 
       \delta(\bar z_0 - 
       \bar z_0^\prime) \epsilon(\dot z_0^\prime) \dot{\bar z} _0
       {(\bar z^\mu - \bar z^{\prime \mu}) \dot{\bar z} ^{\prime \nu} -
      (\bar z ^\nu - \bar z^{\prime \nu})\dot {\bar z} ^{\prime \mu} 
      \over | \bar {\vec z} - \bar{\vec z}^\prime| 
      \sqrt {\dot {\bar z_0}^{\prime 2} - 
      \dot {\bar {\vec z}}_{\rm T}^{\prime 2}}} 
\end{array} 
\end{equation}

   Notice also

\begin{equation}
\begin{array}{l}
{\delta^2 \over \delta S^{\rho \sigma}(z(\tau^\prime))
       \delta S^{\mu \nu}(z(\tau))} E_{\rm pert}(\bar z - \bar z^\prime, 
        \dot{\bar z}, \dot{\bar z} ^\prime )= \nonumber \\
\qquad \qquad = - 8\pi {4 \over 3} \alpha_{\rm s} 
        \big [\delta (\tau - \bar \tau)
        \delta( \tau^\prime - \bar \tau ^\prime)
       (\delta_\mu^\alpha \partial_\nu - \delta_\nu^\alpha \partial_\mu)
       (\delta_\rho^\beta \partial_\sigma 
       - \delta_\sigma^\beta \partial_\rho) 
       D_{\alpha \beta}(\bar z - \bar z ^\prime)  
\end{array}
\end{equation}

\noindent
and

\begin{equation}
{\delta^2 \over \delta S^{\rho \sigma}(z(\tau^\prime))
       \delta S^{\mu \nu}(z(\tau))} E_{\rm conf} 
       (\bar z - \bar z^\prime, 
        \dot{\bar z}, \dot{\bar z} ^\prime ) = 0 \, .
\end{equation}

\noindent
Then

\begin{eqnarray}{l}
{ \cal S} _0^s E(z - z^\prime , \dot z, \dot z ^\prime)  
        ({\cal S}_0^s)^{-1}=
        E-{1 \over 4} \sigma^{\mu \nu} 
        \int_0^s d \theta \bigg [{\delta
        \over \delta S^{\mu \nu}(\zeta (\theta))} ,
        \, E \bigg] + \nonumber \\
\qquad \qquad \qquad + {1 \over 32} \sigma^{\mu \nu}\sigma^{\rho \sigma} 
        \int_0^s d \theta  \int_0^\theta d \theta^\prime \bigg [{\delta
        \over \delta S^{\rho \sigma}(\zeta (\theta^\prime))} ,
        \,\bigg [{\delta
        \over \delta S^{\mu \nu}(\zeta (\theta))} , \, E 
        \bigg] \bigg] + \dots
        = \nonumber \\ 
= E(z - z^\prime , \dot z, \dot z ^\prime)-{1 \over 4}
       \sigma^{\mu \nu} \bigg \{
        4\pi {4 \over 3} \alpha_{\rm s} 
       (\delta_\mu^\alpha \partial_\nu 
        - \delta_\nu^\alpha \partial_\mu ) 
       D_{\alpha \beta}(z - z ^\prime) \dot z^{\prime \beta} 
       + \nonumber \\
\qquad  \qquad \qquad + \sigma  \delta(z_0 - z_0^\prime) 
      \epsilon(\dot z_0) \dot z _0^\prime
       {(z^\mu - z^{\prime \mu}) \dot z ^\nu -
      (z ^\nu - z^{\prime \nu})\dot z ^\mu 
      \over |\vec z - \vec z^\prime| 
      \sqrt {\dot z_0^2 - 
      \dot {\vec z}_{\rm T}^ 2}}- \nonumber \\
\qquad \qquad \qquad - 4\pi {4 \over 3} \alpha_{\rm s}\
       (\delta_\mu^\beta \partial_\nu -
        \delta_\nu^\beta \partial_\mu) D_{\alpha \beta}(z 
      - z ^\prime) \dot z ^\alpha  - \nonumber \\
\qquad \qquad \qquad - \sigma  \delta(z_0 - z_0^\prime) 
       \epsilon(\dot z_0^\prime) \dot z _0
       {(z^\mu - z^{\prime \mu}) \dot z ^{\prime \nu} -
      (z ^\nu - z^{\prime \nu})\dot z ^{\prime \mu} 
    \over | \vec z - \vec z^\prime| 
      \sqrt {\dot {\bar z}_0^{\prime 2} - 
      \dot {\vec z}_{\rm T}^{\prime 2}}} \bigg \} - \nonumber \\
\qquad \qquad \qquad - {1\over 16} 4\pi {4 \over 3} \alpha_{\rm s} 
       \sigma^{\mu \nu} \sigma^{\rho \sigma}
     (\delta_\mu^\alpha \partial_\nu - \delta_\nu^\alpha \partial_\mu)
       (\delta_\rho^\beta \partial_\sigma 
       - \delta_\sigma^\beta \partial_\rho) 
       D_{\alpha \beta}(z - z ^\prime)  
\label{eq:A8}
\end{eqnarray} 

   Finally, after the Legendre transformation, ${\dot z}^\mu$ and 
$\dot z{\prime \mu}$ are simply replaced in (\ref{eq:A8}) by 
$p ^\mu$ and $p^{\prime \mu}$.


\section{Explicit expression of the sd kernels}

\indent
    Evaluating explicitly the trace in Eq. (\ref{eq:kernexpl}), one obtains

\begin{equation}
\left \{
\begin{array}{l}
R_{00}(k,l)= - \hat I_{0;0} - 2 g^{\mu \rho} g^{\nu \sigma} 
           \hat I_{\mu \nu ;
            \rho \sigma} \nonumber \\
R_{11}(k,l)=\hat I_{0 ;0} - 8 g^{\nu \sigma} \hat I_{0 \nu ; 0 \sigma} 
        + 2  g^{\mu \rho} g^{\nu \sigma} \hat I_{\mu \nu ; \rho \sigma} 
        \nonumber \\
R_{12}(k,l) = - R_{21}(l,k) = 4i\hat I_{0;0i} \hat l^i - 
         8 g^{\nu \sigma} \hat I_{0 \nu ; i \sigma} \hat l^i  \nonumber \\
R_{22}(k,l) = \hat I_{0;0}\hat k \cdot \hat l 
          + 8 g^{\nu \sigma} \hat I_{i \nu ;j \sigma} \hat k^i \hat l^j
         + 2  g^{\mu \rho} g^{\nu \sigma} \hat I_{\mu \nu ; \rho \sigma} 
         \hat k \cdot \hat l
         + 2i(\hat I_{0;ij}-\hat I_{ij;0} \hat k^i \hat l^j   \nonumber \\
R_{33}(k,l) = \hat I_{0;0}-8(g^{ij} +2 \hat k^i \hat l^j) \hat I_{0i;0j}
         + 4 g^{\nu \sigma} \hat I_{i \nu ;j \sigma} \hat k^i \hat l^j 
         + 2 g^{\mu \rho} g^{\nu \sigma} \hat I_{\mu \nu ; \rho \sigma}
\end{array} \right.
\label{eq:B1}
\end{equation}

\noindent
where the arguments are intended as in (\ref{eq:sdexpl}).


\section{Euclidean form of the confining kernels }

\indent
   The euclidean form of $\hat R_{rs}(k,l)$ is obtained from 
Eq. (\ref{eq:B1}) and (\ref{eq:imom}) by setting
$p=p^\prime = q$ and then making the substituitions $Q_0 \to iQ_4, \ q_0 
\to iq_4, \ R_{12} \to iR_{12}, \ R_{21} \to -iR_{21}$, according 
to the rules given in Sec. IV.

    We report the explicit expression for the confinement parts, 
the only non trivial:

\begin{equation}
\left \{
\begin{array} {l}
R_{00}^{\rm conf}(k,l)= \sigma \int d^3 \vec \zeta 
       e^{-i \vec Q \cdot \vec \zeta} |\vec \zeta |\, |q_4|\sqrt{q_4^2 +
       \vec q_{\rm T}^2 } 
        \nonumber \\
R_{11}^{\rm conf}(k,l)= - \sigma \int d^3 \vec \zeta 
       e^{-i \vec Q \cdot \vec \zeta} |\vec \zeta |\, |q_4|\sqrt{q_4^2 +
       \vec q_{\rm T}^2 }
        \nonumber \\
R_{12}^{\rm conf}(k,l) = R_{21}^{\rm conf}(l,k) 
       =-4i \sigma \int d^3 \vec \zeta
       e^{-i \vec Q \cdot \vec \zeta} 
        \epsilon (q_4){q_4^2 \over |\vec \zeta |}
        \, { \vec \zeta \cdot \hat l \over \sqrt{q_4^2 
        + \vec q_{\rm T}^2}} 
        \nonumber \\
R_{22}^{\rm conf}(k,l) = - \sigma \int d^3 \vec \zeta 
       e^{-i \vec Q \cdot \vec \zeta}|q_4| [|\vec \zeta |\, 
       \sqrt{q_4^2 +
       \vec q_{\rm T}^2 } + 4i 
        \, { (\vec \zeta \cdot \hat k)(\vec q \cdot \hat l)
       -(\vec \zeta \cdot \hat l)(\vec q \cdot \hat k) 
       \over  |\vec \zeta |
       \sqrt{q_4^2 + \vec q_{\rm T}^2}}]
        \nonumber \\
R_{33}^{\rm conf}(k,l) =- \sigma \int d^3 \vec \zeta 
       e^{-i \vec Q \cdot \vec \zeta} |\vec \zeta |\, 
      |q_4|\sqrt{q_4^2 +
       \vec q_{\rm T}^2 } 
\end{array} \right.
\label{eq:C1}
\end{equation}


\begin{figure}
\caption{Wilson loop $\Gamma_{q \bar q}$ for the four points Green 
 function formed by the quark and the 
 antiquark trajectories (full lines) closed by two Swinger strings (broken 
 lines).}
\label{fig1}
\end{figure}

\begin{figure}
\caption{Wilson loop $\Gamma_q$ for the quark propagator formed by the 
 quark trajectory closed by a Scwinger string.}
\label{fig2}
\end{figure}

\begin{figure}
\caption{Loop $\Gamma_{q \bar q}$ for conciding end points with the quark
 trajectory going also bakwards in time.}
\label{fig3}
\end{figure}

\end{document}